\documentclass[pre,10pt,twocolumn,superscriptaddress,floatfix,noeprint]{revtex4-1}

\usepackage{amsmath,bm}
\usepackage{latexsym}
\usepackage{amssymb}
\usepackage{graphics,epstopdf}
\usepackage{hyperref}
\usepackage[caption=false]{subfig}
\usepackage[percent]{overpic}
\usepackage{comment}

\usepackage{braket}
\hypersetup{
	colorlinks = true,
	linkcolor =blue,
	citecolor=blue, 
	urlcolor=blue 
}

\usepackage{mathtools}

\newcommand{\Tr}[2]{\ensuremath{\mathrm{Tr}_{#2}\left[#1\right]}}
\newcommand{\avg}[1]{\ensuremath{\langle #1 \rangle}}

\MakeRobust{\eqref}
\newcommand{\eqnref}[1]{Eq.~\eqref{#1}}
\newcommand{\figref}[1]{Fig.~\eqref{#1}}

\newcommand{\hrho}{\hat{\rho}}
\newcommand{\Hop}{\hat{H}}

\newcommand{\Uop}{\hat{U}}
\newcommand{\Pop}{\hat{P}}
\newcommand{\Xop}{\hat{X}}
\newcommand{\Vop}{\hat{V}}

\newcommand{\commu}[2]{\ensuremath{\left[#1,#2\right]}}

\begin{document}
\title{Controlling Work Output and Coherence in Finite Time Quantum Otto Engines Through Monitoring}
\author{Rahul Shastri}
\email{shastri.rahul@iitgn.ac.in}
\affiliation{Indian Institute of Technology Gandhinagar, Palaj, Gujarat 382055, India}
\author{B. Prasanna Venkatesh}
\email{prasanna.b@iitgn.ac.in}
\affiliation{Indian Institute of Technology Gandhinagar, Palaj, Gujarat 382055, India}
\begin{abstract}
We examine the role of diagnostic quantum measurements on the work statistics of a finite-time quantum Otto heat engine operated in the steady-state. We consider three pointer-based measurement schemes that differ in the number of system-pointer interactions and pointer measurements. We show that the coherence of the working substance and the work output of the engine can be controlled by tuning the monitoring measurements. Moreover, for a working substance consisting of a two-level system we show that while all three schemes reproduce the predictions of the cycle without any monitoring for the average work in the limit of infinitely weak measurement, only two of the schemes can reproduce the two-point projective measurement results in the limit of strong measurement. 
%These schemes interpolate between the two extreme limits of completely unmonitored heat engines (energetics determined by the average internal energy change of the working substance) and Otto engine with the two-point projective (TPM) measurement (energetics determined by outcomes of the projective energy measurement at each point of the Otto cycle). In particular we consider situations where the diagnostic measurement is made by measuring the pointer at each point of the cycle  (scheme $\mathrm{S1}$),  after the completion of each stroke (scheme $\mathrm{S2}$), and only after the completion of the full cycle (scheme $\mathrm{S3}$). These schemes differ in when and how energy measurements are made during the Otto cycle and are hence shown to have varying degrees of coherence retention throughout the Otto cycle.%

\end{abstract}

\maketitle

\section{\label{sec:Introduction}Introduction}

%The study of Quantum heat engines (QHE) has always been of great importance both technologically and for a theoretical understanding of the field of quantum thermodynamics. The first study of QHEs goes back to the pioneering works \cite{Basov1955,Alicki1979}. Since then, the quantum heat engine has been studied extensively. One of the central questions of these studies is whether genuine quantum effects can give an advantage over the classical counterpart in terms of average work output and efficiency \cite{Myers2022,Cangemi2023,Bhattacharjee2021}. Not only the first moments of work(heat) but more recently people have also given attention to the study of work(heat) fluctuations and statistics \cite{Denzler2021,Shastri2022,Saryal2021,Mohanta2022,Gramajo2023,Jaseem2022,DPoletti}.  
A central question in the study of quantum heat engines (QHEs) \cite{Kosloff2013,Sai2016,Friedenberger_2017,Alicki2018,Deffner2019,Mitchison,Bhattacharjee2021,Myers2022,Cangemi2023} %\pvcheck{cite reviews on QHE, be liberal cite all the big ones + one chapter on what is quantum about heat engines from Alicki in Felix Binder's book}% 
is whether genuine quantum effects can give an advantage over classical counterparts in terms of performance metrics like average work output and efficiency. In this context, quantum coherence in the energy basis of the working system is a genuine quantum resource that can be expected to play a role in distinguishing QHEs \cite{Latune2021}. Indeed, it has been shown that an unconventional bath containing quantum coherence can enhance the engine efficiency beyond Carnot \cite{Scully2003} in a model of continuous QHEs. On the other hand, for cyclic QHEs it was initially thought that quantum coherence in the working substance generated through the non-quasistatic driving is detrimental to the performance of the quantum heat engine and was termed as quantum friction 
%(\pvcheck{add also other papers on Quantum Friction + Otto})%
\cite{DelCampo2016,Brandner2017,delCampo2018,Insinga2020,Cakmak2020}. Though, in some recent works it has been observed that in certain cases quantum coherence can be advantageous and has been termed as quantum lubrication \cite{Latune2021,Camati2019,Dann2020a,Aimet2023,Kosloff2006}.
%\pvcheck{I included Aimet paper into the quantum lubrication type study - if you think there is a genuine difference we can keep the next line}
%Moreover, recently it has also been shown the possibility of converting quantum coherence directly into useful work \cite{Aimet2023}. 
%Note that quantum coherence is a basis dependent property and in our discussions, quantum coherence refers to coherence in an energy basis. 
%Coming back to the genuine quantum effects, 

In early discussions of cyclic QHEs, typically work and heat exchange was discussed at the level of averages by examining the internal energy changes in the strokes \cite{Kosloff2013,Quan2007}. More recently, there has been a sustained interest in exploring the fluctuations as well as the statistical distribution of work and heat in cyclic QHEs \cite{Campisi_2015Fluctuations,DPoletti,Funo2018,Landi2019TUR,Marti2021TUR,Mitchison2021Fluctuations,Denzler2021,Saryal2021,Jaseem2022,PP2022,Shastri2022,Mohanta2022,Gramajo2023} %\pvcheck{Order these references/cite them by year of publication. It is not nice to cite our paper early and Landi's TUR later. Also Check All references by formatting the bibtex carefully. Many references are not properly formatted.}%. 
%\pvcheck{Add some rep TUR papers also, for instance the one by Landi and Goold..ensure all cyclic QHE papers with fluctuations by the big guys are cited.}% 
However, to study the work and heat statistics for a quantum heat engine, one has to also specify the monitoring or diagnostic scheme \cite{Lutz2007WorkFluctuation,TalknerRMP2011,Talkner2016} used to measure the thermodynamic variables. Invariably this monitoring will affect the underlying work and heat statistics due to the measurement back-action on the system \cite{Perarnau-Llobet2017}. 

The most widely used monitoring scheme to evaluate work/heat statistics is the two-point projective energy measurement scheme (TPM) \cite{TalknerRMP2011,Lutz2007WorkFluctuation}.
%(\pvcheck{Peter+Eric Lutz original paper, no need to cite any one paper like Denzler here I think as all papers we have cited before on fluctuations + QHE use TPM})% 
A key disadvantage of using TPM scheme for defining work fluctuations is that it destroys any quantum coherence present in the system state due to the strong measurement back-action. Thus the invasive TPM scheme cannot capture the effect of quantum coherence on work fluctuations \cite{Perarnau-Llobet2017,Talkner2016}. Some strategies that have been employed to mitigate this issue is to replace the projective measurements of TPM scheme with the weak energy measurements \cite{Talkner2016,Roncaglia2014,DeChiara2018} 
%(\pvcheck{add paper by Gabriele Dechiara and Roncaglia})
or propose different schemes to define work distribution for a quantum system containing coherence \cite{Solinas2015WorkDist,Lostaglio2018Quasidistribution,Paternostro2020Quasiditributio,Rodrigues2023}.
%(\pvcheck{cite quasiprobability people here the Finnish guy especially})%
In particular, for the case of Quantum Otto Engines (QOE) with multiple cycles, the effect of monitoring on the performance of quantum heat engines has been considered both including and without outcoupling \cite{Son2021,Son2022}. There the diagnostic measurements were implemented using a pointer interacting with the working system and different schemes of measurement realised by varying the number of system pointer interactions were considered. Remarkably, it was shown that monitoring after some number of engine cycles outperforms the scheme where the output is monitored in every stroke in terms of average work output and reliability of work output (relative fluctuations of work) \cite{Son2021}.

In this paper we uncover the impact of monitoring quantum measurements on the work output and coherences in the \emph{steady state} operation (to be defined exactly below) and work output of a finite time QOE operated over one cycle. In line with \cite{Roncaglia2014,Thingna2020,Son2021,Son2022} we model the measurements by coupling the working substance to a pointer system. Moreover, we consider three distinct schemes that differ in the number of system-pointer interactions and hence in the extent of back-action. We show that in all three distinct schemes, we can control the extent of coherence retention by tuning the pointer system to vary the measurement back-action strength. While all three schemes reproduce the predictions of the unmonitored QOE cycle for the average work in the limit of infinitely weak measurement, only two of the schemes can reproduce the TPM result in the limit of strong measurement for the specific choice of working system. We note that our work is complementary but distinct from \cite{Son2021,Son2022} which was concerned with a QOE run over many cycles. In a sense our work is concerned only with the asymptotic limit of the QOE after many cycles.
%To better understand the effect of monitoring on the coherence present in the quantum mechanical working substance and hence on the performance of the QHE, here we consider a finite time quantum Otto engine cycle setup with three monitoring schemes of varying degrees of measurement back-action. Here we are interested in the steady-state  operation of the Otto cycle as we will define later in the section. This is in contrast with the work done in \cite{Son2021}. In their work, the authors were more interested in the transient regime, and their main aim was to compare the work and heat statistics for the two monitoring schemes as a function of a number of cycles. This analysis will help in better understanding the role of monitoring on the performance of the quantum heat engines. 

The article is structured as follows. We begin by introducing the setup for a finite time Otto cycle consisting of a general quantum mechanical working substance in Subsec.~\ref{subsec:FTQOC}, followed by the description of the unmonitored QOE cycle in Subsec.~\ref{subsec:Unmeas} and the pointer-system coupling paradigm to monitor the system in Subsec.~\ref{subsec:PointerMeas}. Subsequently in Subsecs.~\ref{subsec:Scheme 1},\ref{subsec:Scheme 2}, \ref{subsec:Scheme 3} we describe in detail the three different measurement schemes and present formal expressions for the associated work and heat distributions. In Sec.~\ref{sec:Results} we illustrate our results by considering a QOE with a two-level system (TLS) working substance (see Subsec.~\ref{subsec:System}). In particular we obtain analytical results for the restricted scenario of a perfectly thermalizing isochoric stroke in Subsec.~\ref{subsec:Perfectly Thermalising Cooling Isochore} and present numerical results in the more general protocols with a TLS working substance in Subsec.~\ref{subsec:Numerics}. We summarize our results and conclude in Sec.~\ref{sec:Conclusion}. %~\ref{subsec:FTQOC}, we introduce  and briefly discuss the issue of strong measurement back-action of TPM scheme. In Sec.~\ref{subsec:Scheme 1}, Sec.~\ref{subsec:Scheme 2} and Sec.~\ref{subsec:Scheme 3} we introduce the three measurement schemes. In Sec.~\ref{subsec:System}, we describe the setup for the Otto cycle consisting of a two-level system (TLS) as a working substance. Then in the rest of the Sec.~\ref{sec:Results}, we discuss the main results. Then we summarise the findings and conclude in Sec.~\ref{sec:Conclusion}
\section{\label{sec:Model} Setup \& Measurement Schemes}
\subsection{\label{subsec:FTQOC} Finite Time Quantum Otto Cycle}
We consider a four-stroke finite-time Quantum Otto engine (QOE) cycle consisting of a quantum mechanical working substance described by a time-dependent Hamiltonian $\Hop(t)$. The four consecutive strokes of the Otto cycles are compression work stroke ($1\rightarrow 2$), dissipative hot isochore ($2\rightarrow 3$), expansion work stroke ($3\rightarrow 4$), and dissipative cold isochore ($4\rightarrow 1$). Briefly, the system starts with the initial state $\hrho_1$ and the compression work stroke ($1\rightarrow 2$) is carried out, while the system is isolated from the baths, by changing a time-dependent control parameter $\lambda(t)$ that determines the working substance hamiltonian. This can be described by unitary time evolution operation $\hrho_2=\Uop_1\hrho_1\Uop_1^{\dagger}$. The expansion work stroke ($3\rightarrow 4$) is carried out by varying the control parameter as $\tilde{\lambda}(t)$ in a time-reversed manner with respect to the compression \footnote{This is typical though see \cite{DPoletti,Mohanta2022,Shastri2022,Makouri2023} for exceptions}. In this case, the time evolution of the state is described by  $\hrho_4=\Uop_2\hrho_3\Uop_2^{\dagger}$, where $\Uop_2$ is the time evolution operator corresponding to the time-reversed protocol $\tilde{\lambda}(t)$. The dissipative hot isochore ($2\rightarrow 3$) and cold isochore ($4\rightarrow 1$) are modeled by completely positive trace preserving (CPTP) quantum maps $\hrho_3=\Phi_{\beta_{h}}(\hrho_2)$ and $\hrho_5=\Phi_{\beta_{c}}(\hrho_4)$ respectively acting on the system density matrix, where $\beta_h$ and $\beta_c$ are the inverse temperatures corresponding to the two heat baths ($\beta_c>\beta_h$). While we keep the description of the thermalization CPTP maps general in this section, in our discussion in Section~\eqref{sec:Results} with a TLS working substance, they will be generated from the time-evolution given by a Gorini-Kossakowski-Sudarshan-Lindblad (GKSL) master equation.

\subsection{Unmonitored Quantum Otto Cycle}
\label{subsec:Unmeas}
The entire QOE cycle, without considering any diagnostic measurements, can be represented as a CPTP map acting on the system's density matrix \cite{Dann2020}
\begin{align}
\Phi_{\mathrm{UM}}(\hrho_1) = \Phi_{\beta_c}\left[\Uop_2\Phi_{\beta_h}\left(\Uop_1 \hrho_1\Uop_1^{\dagger}\right)\Uop_2^{\dagger}\right].
\label{UMOttoCycle}
\end{align}
We will henceforth refer to this as the unmonitored Otto cycle. Note that throughout our study, we will treat each of the Otto cycle strokes as ones with a finite duration. For simplicity, we assume that $\tau_u$ is the duration of each unitary work stroke, $\tau_b$ is the duration of each dissipative stroke, and $\tau=2(\tau_u+\tau_b)$ is the total cycle time. Note that if all the strokes are of finite time, then the system state after one complete cycle $\hrho_5=\Phi_{\mathrm{UM}}(\hrho_1)$ generally differs from the initial state $\hrho_1$ and hence the cycle does not close. This leads to transient behavior during a number of cycles until a steady state is reached \cite{Son2021}. The steady state is the fixed point of the cycle CPTP map Eq.~\eqref{UMOttoCycle} satisfying,
\begin{align}
\hrho^{\mathrm{ss}}_{\mathrm{UM}}=\Phi_{\mathrm{UM}}(\hrho^{\mathrm{ss}}_{\mathrm{UM}}).
\label{UMCPTPmapSS}
\end{align}

In our study, we are specifically interested in the steady-state operation of the Otto cycle and not the transient dynamics over many cycles \cite{Son2021}. The existence of unique invariants of the CPTP map guarantees monotonic
%(\pvcheck{does it have to be monotonic?}-\textbf{This statement is based on \cite{Dann2020}})% 
convergence to the fixed point independent of the initial state and depends only on the system Hamiltonian and the unitary and dissipative strokes \cite{Dann2020}. Given the initial state of the system $\hrho_1$, the CPTP map $\Phi_{\mathrm{UM}}$ completely describes the entire Otto cycle and the final state after the completion of one cycle. The energetics of the unmonitored cycle are described by defining the average total work output and heat exchange by tracking the average energy change of the working system \cite{Dann2020}.

\subsection{Quantum Measurement of Energy using Pointer systems}
\label{subsec:PointerMeas}
Recently, there has been significant interest in going beyond the average work output and heat and studying in general the work and heat statistics of quantum heat engine cycles. A key step in constructing the statistical distributions of process dependent thermodynamic variables like work and heat in quantum systems is the introduction of a diagnostic quantum measurement scheme to read out the relevant energies. In this context the most widely used scheme is the two-point measurement (TPM) scheme \cite{TalknerRMP2011}. In the TPM scheme, the work or heat exchanged by the system is measured by making a projective measurement of the system's energy at the beginning and after the completion of the process. One of the disadvantages of the TPM scheme is that the projective measurement is extremely invasive and kills the coherence (in energy basis) present in the initial state \cite{Perarnau-Llobet2017}. In the context of the QOE cycle, to construct the work and heat distribution, one has to make projective measurements of the energy at four points of the cycle and the back-action on the system from this invariably destroys any coherence present before the measurement. It has been shown that such residual coherence can in fact enhance the engine performance \cite{Camati2019}. Given that coherence can be a thermodynamical resource in this sense, it is desirable to understand how one can control the measurement back-action to preserve coherence and conceivably improve engine performance. One way to reduce the measurement back action is to use a less invasive or weak energy measurement as we describe next. 

Along this line, following \cite{Thingna2020,Son2021,Son2022}, we model a weak energy measurement by first coupling the system to a pointer and then projectively measuring the pointer. Moreover, as shown in \cite{Roncaglia2014,Thingna2020, Son2021,Son2022} this approach can also be tailored to measure the sums and differences of energy over quantum processes and hence work and heat during thermodynamic strokes. Such flexibility to measure different thermodynamic observables such as work during a stroke or sums of work in different strokes is enabled by controlling the contact interactions between the pointer and system as well as when the pointer is measured. Consequently, the extent of measurement back-action can also be controlled in this manner. In what follows, we will discuss three distinct measurement schemes that precisely do this. To set the stage, let us now discuss the description of the pointer system common to all the schemes. The pointer is taken as a free particle in one dimension with Hamiltonian $\bm{\Hop}=\frac{\bm{\Pop}^2}{2M}$ with mass $M$, conjugate position $\bm{\Xop}$ and momentum $\bm{\Pop}$  operators satisfying commutation relation $\commu{\bm{\Xop}}{\bm{\Pop}}=i\hbar \mathrm{I}$. We will use bold fonts for pointer operators and normal fonts for the system operators throughout the article. The interaction between the system and pointer is of the form $\Hop_{I}=\lambda\Hop\otimes\bm{\Pop}$ where $\lambda$ is the interaction strength. %This particular choice of interaction serves two purposes as we will see later (\pvcheck{Make sure you explain the purposes eventually - not very clear}).
The coupling operator's choice ensures that the pointer's position is shifted by an amount proportional to the system's energy. Moreover, since the interaction commutes with the system's hamiltonian, the average energy of the system is unchanged by the pointer coupling. These general features are desirable for any pointer scheme. We assume that before the first interaction with the system, the pointer state is initialized in a pure Gaussian state with zero mean and finite variance $\sigma^2$,
\begin{align}
\bm{\hrho}_{\sigma} &= \int dydy^{\prime} \frac{1}{\sqrt{2\pi\sigma^2}}e^{-\frac{1}{4\sigma^2}(y^2+y^{\prime 2})} \ket{y}\bra{y^{\prime}}.
\label{GaussianState}
\end{align}
%where $ \rho_{\sigma}(y,y^{\prime})=\frac{1}{\sqrt{2\pi\sigma^2}}e^{-\frac{1}{4\sigma^2}(y^2+y^{\prime 2})}$. 
In order to measure the system's energy we turn on the system-pointer interaction for a time $\tau_p$. During this time, the initial product state of the system and pointer $\hrho\otimes\bm{\hrho}_{\sigma}$ evolves to $\Vop(\hrho\otimes\bm{\hrho}_{\sigma})\Vop^{\dagger}$  where the unitary $\Vop=e^{-i\lambda\tau_p\Hop\otimes\bm{\Pop}/\hbar}$. The interaction time $\tau_p$ is taken short enough such that the system's free dynamics is negligible. Following this we make a projective measurement of the position of the pointer. The shift of position of the pointer $x$ after the interaction with the system gives the unbiased value of the energy of the system $x=\tilde{\lambda}E$, with $\tilde{\lambda}=\lambda\tau_p/\hbar$. We set $\hbar=\tilde{\lambda}=1$ in what follows. The pointer measurement results in a non-normalized post-measurement state of the system conditioned on the outcome $x$,
\begin{align}
\Phi_{x}(\hrho)=\Tr{\bm{\Pi}_{x}\Vop(\hrho\otimes\bm{\hrho}_{\sigma})\Vop^{\dagger}\bm{\Pi}_{x}}{p},
\label{OneMeasurementExample}
\end{align}
where $\bm{\Pi}_{x}=\ket{x}\bra{x}$ is the projection operator of the pointer. Here the trace is taken with respect to the pointer Hilbert space.
The probability of the outcome $x$ is obtained by tracing over the system density matrix as
\begin{align}
p(x) = \mathrm{Tr}_s[\Phi_{x}(\hrho)].
\label{OneMeasurementProbDistExample}
\end{align}
We can show from the fact that $\commu{\Hop}{\Hop_{I}}=0$, the expectation value of the system's energy before measurement is given by the average value of the pointer measurement outcomes \emph{i.e.} $\mathrm{Tr}[\hrho\Hop] = \int xdxp(x)$. Simplifying  ~\eqnref{OneMeasurementExample} we get,
\begin{align}
\Phi_{x}(\hrho) &= \sum_{m,m'} \Pi_{m}\hrho\Pi_{m'}e^{-\frac{1}{8\sigma^2}\left(e_{m}-e_{m'}\right)^2}\nonumber\\
&\times G_{\sigma}\left(x-(e_{m}+e_{m'})/2\right),
\label{OneMeasurementExample2}
\end{align}
where $G_{\sigma}(x-x_0)$ is the normalized Gaussian function with variance $\sigma$ and mean $x_0$. $\Pi_{m}=\ket{m}\bra{m}$ is the projection operator corresponding to the system Hamiltonian with energy eigenstate $\ket{m}$ and eigenvalue $e_m$. Let us now examine the post-measurement state in two extreme cases. When the initial pointer width is much smaller than the smallest energy gap (i.e. $\sigma<<\mathrm{min}_{m,m'}|e_m-e_{m'}|$), we have the precise pointer limit where the measurement is able to resolve the individual energies of the system. In an extreme version of this limit with $\sigma \rightarrow 0$, we recover the standard projective energy measurement \emph{i.e.}
\begin{align*}
    \Phi_{x}(\hrho) \stackrel{\sigma \rightarrow 0}{\approx} \Pi_{m}\hrho\Pi_{m},
\end{align*}
with $e_m = x$. In contrast,  when the pointer width is much larger than the largest energy gap (i.e. $\sigma>>\mathrm{max}_{m,m'}|e_m-e_{m'}|$), we obtain the imprecise pointer limit where the measurement cannot resolve the individual energies of the system. In the extreme limit of $\sigma \rightarrow \infty$, we get a completely `undisturbed' post-measurement state: 
\begin{align*}
    \Phi_{x}(\hrho) \stackrel{\sigma \rightarrow \infty}{\approx} \sum_{m,m'} \Pi_{m}\hrho\Pi_{m'} = \hrho,
\end{align*}
and the pointer measurement does not give any information about the energy of the system. In this manner we can control the measurement back-action on the system by tuning the parameter $\sigma$ characterizing the initial state of the pointer.    

In addition to controlling measurement back-action by tuning the pointer state, we can also choose to measure the sums/differences of the system energy (hence work or heat) by multiple interactions and a single delayed measurement as discussed in \cite{Roncaglia2014,Thingna2020, Son2021,Son2022}. In some cases, this further reduces the number of measurements and hence also reduces measurement back-action on the system state. In particular, we will next consider a finite time Otto cycle with the diagnostic measurements made by measuring the pointer at each point of the cycle, after the completion of each stroke, and only after the completion of the full cycle. We will demonstrate that these schemes have varying degrees of coherence retention and hence work output and fluctuations. %In the next section, we will describe three measurement schemes in detail and then compare the work and heat statistics for these measurement schemes.Motivated by this, in what follows, we present three different measurement schemes. 
%%%%%%%%%%%%%%%%%%%%%%%%%%%%%%%%%%
\begin{figure*}
	\centering
\includegraphics[scale=1.0]{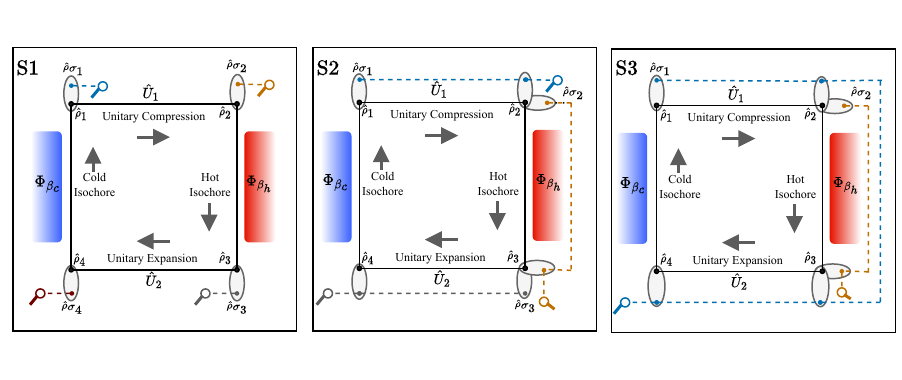}
    \caption{ Schematic of the diagnostic measurement schemes $\mathrm{S1}$, $\mathrm{S2}$ and $\mathrm{S3}$ (from left to right). The working system and pointer system state at the four points of the cycle are represented by $\rho_i$ and $\rho_{\sigma_i}$ respectively ($i=1,2,3,4$).  Pointer measurement (denoted by magnifying glass symbol) is carried out at four points of the cycle for $\mathrm{S1}$, at points $2$ and $3$ for $\mathrm{S2}$ and at points $3$ and $4$ for $\mathrm{S3}$. }
\label{fig:MeasurementSchemeSchematic}
\end{figure*}
%%%%%%%%%%%%%%%%%%%%%
\subsection{\label{subsec:Scheme 1} Measurement Scheme 1}
In the first scheme, denoted $\mathrm{S1}$, the work and heat statistics for the Otto cycle are constructed by measurements of the energy of the system at four points of the Otto cycle. This requires four (distinct) pointers, each initialized in a Gaussian state of width $\sigma_k$ and four measurements at the four end-points of the cycle denoted by $k=1,2,3,4$ (see schematic Fig.~(\ref{fig:MeasurementSchemeSchematic}a). The energy measurements are carried out as described in the previous sub-section. Let us denote by $\Hop_k=\sum_{m=1}^{d}e_{m}^{(k)}\Pi_{m}^{(k)}$ the Hamiltonian of the system at the different points with $\Pi_{m}^{(k)}$ the projection operator corresponding to the system energy eigenstates $\ket{e_m^{(k)}}$ with eigenvalues $e_m^{(k)}$. Moreover, in the notation of our finite time QOE, we have that $\Hop(t_1) = \Hop_1 = \Hop(t_1+2 \tau_u + \tau_b) = \Hop_4$  and $\Hop(t_1+\tau_u) = \Hop_2 = \Hop(t_1+\tau_u+\tau_b) = \Hop_3$. The non-normalized state of the working substance system after one cycle conditioned on the pointer measurement outcomes of $\vec{x}=(x_1,x_2,x_3,x_4)$ reads \cite{Son2021,Son2022}
\begin{align}
\Phi_{\mathrm{S1},\Vec{x}}(\hrho_1) &= \sum_{\Vec{m},\Vec{m}'} \mathcal{S}^{\Vec{m},\Vec{m}'}(\hrho_1)\Lambda_{\mathrm{S1}}^{\vec{m},\vec{m}'}\nonumber\\
&\times\prod_{k=1}^{4}G_{\sigma_k}(x_k-(e_{m_k}^{(k)}+e_{m_k'}^{(k)})/2),
\label{non-normalisedSystemState2}
\end{align}
% \begin{align}
% \Phi_{\mathrm{S1},\Vec{x}}(\hrho_1) &= \sum_{\Vec{m},\Vec{m}'} \mathcal{S}^{\Vec{m},\Vec{m}'}(\hrho_1) \prod_{k=1}^{4}\rho_{\sigma_k}\left(x_k-e_{m_k}^{(k)},x_k-e_{m_k'}^{(k)}\right),
% \label{non-normalisedSystemState}
% \end{align}
where,
\begin{widetext}
	\begin{align}
\mathcal{S}^{\Vec{m},\Vec{m}'}(\hrho_1)=\Phi_{\beta_c}\left[\Pi_{m_4}^{(1)}U_2\Pi_{m_3}^{(2)}\Phi_{\beta_h}\left(\Pi_{m_2'}^{(2)}U_1\Pi_{m_1}^{(1)}\hrho_1\Pi_{m_1'}^{(1)}U_1^{\dagger}\Pi_{m_2'}^{(2)}\right)\Pi_{m_3'}^{(2)}U_2^{\dagger}\Pi_{m_4'}^{(1)}\right],
\end{align}
\end{widetext}
with $\vec{m}=(m_1,m_2,m_3,m_4),\vec{m}'=(m_1',m_2',m_3',m_4')$ denote the vector of eigenvalues of the respective Hamiltonians $\Hop_k$ at four points of the Otto cycle and
\begin{align*}
\Lambda_{\mathrm{S1}}^{\vec{m},\vec{m}'}=\prod_{k=1}^{4} e^{-\frac{1}{8\sigma_k^2}\left(e_{m_k}^{(k)}-e_{m_k'}^{(k)}\right)^2}.
\end{align*}
%and the sum is over all the possible eigenvalues of the respective Hamiltonians $\Hop_k$.  . Simplifying \eqnref{non-normalisedSystemState} we get,
% \begin{align}
% \Phi_{\mathrm{S1},\Vec{x}}(\hrho_1) &= \sum_{\Vec{m},\Vec{m}'} \mathcal{S}^{\Vec{m},\Vec{m}'}(\hrho_1)\Lambda_{\mathrm{S1}}^{\vec{m},\vec{m}'}\nonumber\\
% &\times\prod_{k=1}^{4}G_{\sigma_k}(x_k-(e_{m_k}^{(k)}+e_{m_k'}^{(k)})/2),
% \label{non-normalisedSystemState2}
% \end{align}
The normalized state can be obtained by integrating out the pointer variables as
\begin{align}
\Phi_{\mathrm{S1}}(\hrho_1) &= \sum_{\Vec{m},\Vec{m}'} \mathcal{S}^{\Vec{m},\Vec{m}'}(\hrho_1) \Lambda_{\mathrm{S1}}^{\vec{m},\vec{m}'}.
\label{scheme1CPTPmap}
\end{align} 
Here $\Phi_{\mathrm{S1}}$ is the CPTP map representing the full thermodynamic cycle with measurements according to scheme S1 acting only on the density matrix of the system. As discussed before the steady-state that we are interested in, $\hrho^{\mathrm{ss}}_{\mathrm{S1}}$, is the fixed point of the cycle CPTP map \eqnref{scheme1CPTPmap} i.e.
\begin{align}
\hrho^{\mathrm{ss}}_{\mathrm{S1}}=\Phi_{\mathrm{S1}}(\hrho^{\mathrm{ss}}_{\mathrm{S1}}).
\label{scheme1CPTPmapSS}
\end{align}
 
In the steady-state operation we evaluate the joint probabality of total work $w$ and heat exchanged with the hot bath $q_h$ as,
\begin{align}
p_{\mathrm{S1}}(w,q_h) &= \int \prod_{k}^{4}dx_k \Tr{\Phi_{\mathrm{S1},\Vec{x}}(\hrho^{\mathrm{ss}}_{1})}{s}\delta[q_h-(x_3-x_2)]\nonumber\\
&\times\delta[w-(x_4-x_3)-(x_2-x_1)]  ,
\label{scheme1ProbDist}
\end{align} 
One can equivalently compute the characteristic function $\chi_{\mathrm{S1}}(k_1,k_2) = \int dwdq_h p_{\mathrm{S1}}(w,q_h)e^{ik_1w}e^{ik_2q_h}$ for the joint work and heat distribution which in this case can be evaluated as,

 % \begin{align}
 % \chi_{\mathrm{S1}}(k_1,k_2)&=\sum_{\Vec{m},\Vec{m}'} \Tr{\mathcal{S}^{\Vec{m},\Vec{m}'}(\hrho^{\mathrm{ss}}_{\mathrm{S1}}) }{s}  \Lambda_{\mathrm{S1}}^{\vec{m},\vec{m}'} e^{i\left(k_1w^{\vec{m},\vec{m}'}+k_2q_h^{\vec{m},\vec{m}'}\right)/2}\nonumber\\
 % &\times e^{-\frac{k_1^2}{2}(\sigma_1^2+\sigma_2^2+\sigma_3^2+\sigma_4^2)-\frac{k_2^2}{2}(\sigma_2^2+\sigma_3^2)+k_1k_2(\sigma_2^2+\sigma_3^2)},
 % \label{scheme1Characteristicfunc}
 % \end{align}%
 \begin{align}
 &\chi_{\mathrm{S1}}(k_1,k_2)=\sum_{\Vec{m},\Vec{m}'}   \Lambda_{\mathrm{S1}}^{\vec{m},\vec{m}'} e^{i\frac{k_1}{2}w^{\vec{m},\vec{m}'}+i\frac{k_2}{2}q_h^{\vec{m},\vec{m}'}-\frac{k_1^2}{2}\sum_{i=1}^4\sigma_i^2}\nonumber \\
 &\Tr{\mathcal{S}^{\Vec{m},\Vec{m}'}(\hrho^{\mathrm{ss}}_{\mathrm{S1}}) }{s} e^{-\frac{k_2^2}{2}(\sigma_2^2+\sigma_3^2)+k_1k_2(\sigma_2^2+\sigma_3^2)},
 \label{scheme1Characteristicfunc}
 \end{align} 
where, $w^{\vec{m},\vec{m}'}=\sum_{k=1}^{4}(-1)^k(e_{m_k}^{(k)}+e_{m_k'}^{(k)})$ and  $q_h^{\vec{m},\vec{m}'}=(e_{m_3}^{(2)}+e_{m_3'}^{(2)})-(e_{m_2}^{(2)}+e_{m_2'}^{(2)})$. From the characteristic function one can obtain moments using $\avg{w^nq_h^m}= \frac{\partial^n\partial^m}{\partial(ik_1)^n\partial(ik_2)^m}\chi_{\mathrm{S1}}(k_1,k_2)|_{\{k_1=0,k_2=0\}}$.
From the expression \eqnref{scheme1Characteristicfunc} we can write down a simplified expression for the joint work and heat distribution function as
\begin{align}
p_{\mathrm{S1}}(w,q_h)&=\sum_{\Vec{m},\Vec{m}'} \Tr{\mathcal{S}^{\Vec{m},\Vec{m}'}(\hrho^{\mathrm{ss}}_{\mathrm{S1}}) }{s}  \Lambda_{\mathrm{S1}}^{\vec{m},\vec{m}'} \nonumber\\
&\times G_{\mathcal{C}}(w-w^{\vec{m},\vec{m}'}/2,q_h-q_h^{\vec{m},\vec{m}'}/2),
\label{scheme1ProbDist2}
\end{align} 
where $G_{\mathcal{C}}(w-w^{\vec{m},\vec{m}'}/2,q_h-q_h^{\vec{m},\vec{m}'}/2)$ is a two-variable Gaussian function with mean values at $\avg{w}=w^{\vec{m},\vec{m}'}/2$ and $\avg{q_h}=q_h^{\vec{m},\vec{m}'}/2$ and covariance matrix 
\begin{align*}
\mathcal{C} &= \begin{pmatrix}
\sigma_1^2+\sigma_2^2+\sigma_3^2+\sigma_4^2 & -(\sigma_2^2+\sigma_3^2)\\
-(\sigma_2^2+\sigma_3^2) & (\sigma_2^2+\sigma_3^2)
\end{pmatrix}.
%\label{scheme1ProbDistCovarianceMat}
\end{align*}
From \eqnref{scheme1ProbDist2} we see that the joint work-heat distribution is sum of Gaussians centered around work values $w^{\vec{m},\vec{m}'}$ and heat values $q_h^{\vec{m},\vec{m}'}$. 
In the limit $\sigma_k \rightarrow 0, \forall k$, each of the above measurements becomes a projective energy measurement and we recover the TPM scheme. This can be explicitly seen by looking at \eqnref{scheme1CPTPmap} where $\Lambda_{\mathrm{S1}}^{\vec{m},\vec{m}'} \stackrel{\sigma_k \to 0}{\longrightarrow} 0$ for all terms with $\vec{m}\neq\vec{m'}$ and the corresponding CPTP map becomes
\begin{align}
\Phi_{\mathrm{TPM}}(\hrho_1) &= \sum_{\Vec{m}} \mathcal{S}^{\Vec{m},\Vec{m}}(\hrho_1).
\label{TPMCPTPmap}
\end{align} 
Consequently, the steady-state density matrix for the TPM scheme is defined by
\begin{align}
\hrho^{\mathrm{ss}}_{\mathrm{TPM}}=\Phi_{\mathrm{TPM}}(\hrho^{\mathrm{ss}}_{\mathrm{TPM}}).
\label{TPMCPTPmapSS}
\end{align}
Taking the $\sigma_k \rightarrow 0$ limit in \eqnref{scheme1Characteristicfunc}, the characteristic function for the TPM case is given by
\begin{align}
\chi_{\mathrm{TPM}}(k_1,k_2)&=\sum_{\Vec{m}} \Tr{\mathcal{S}^{\Vec{m},\Vec{m}}(\hrho^{\mathrm{ss}}_{\mathrm{TPM}}) }{s} \nonumber \\
&\times e^{i\frac{k_1}{2}w^{\vec{m},\vec{m}}+i\frac{k_2}{2}q_h^{\vec{m},\vec{m}}} \label{TPMCharacteristicfunc}
\end{align} % 
The corresponding joint work and heat distribution becomes,
\begin{align}
p_{\mathrm{TPM}}(w,q_h)&=\sum_{\Vec{m}} \Tr{\mathcal{S}^{\Vec{m},\Vec{m}}(\hrho^{\mathrm{ss}}_{\mathrm{TPM}}) }{s}   \nonumber\\
&\times \delta(w-w^{\vec{m},\vec{m}})\delta(q_h-q_h^{\vec{m},\vec{m}}).
\label{TPMProbDist}
\end{align} 
In contrast, for the extreme version of the completely imprecise pointer limit, we get $ \Lambda_{\mathrm{S1}}^{\vec{m},\vec{m}'} \stackrel{\sigma_k \rightarrow \infty}{\longrightarrow} 1$ for all $\vec{m}$ and $\vec{m'}$ and the corresponding CPTP map and steady state become that of the unmonitored Otto cycle \eqnref{UMOttoCycle}. In fact, as we will see in more detail with explicit examples in a later section, once $\sigma_k\gg\mathrm{max}_{i,j}|e_{m_i}^{(k)}-e_{m_j}^{(k)}|$, the probability distributions becomes broad with a very large variance such that it does not give any useful information about the work (heat) statistics beyond the mean values.

\subsection{\label{subsec:Scheme 2} Measurement Scheme 2}
In the second measurement scheme S2, instead of measuring the energies of the system we now measure the work done during compression, $w_1$, and expansion stroke, $w_3$, and the heat exchanged $q_h$ during the 
hot isochore. This requires three pointers and measurements for measuring $w_1$, $q_h$, and $w_3$ respectively, as opposed to $\mathrm{S1}$ which required four pointers and measurements. For instance, as shown schematically in Fig.~(\ref{fig:MeasurementSchemeSchematic}b), to measure the work during compression stroke $w_1$ \cite{Roncaglia2014,Son2021,Son2022}, we first connect the system with the pointer system $1$ with interaction unitary $\Vop_1=e^{-i\Hop_1\otimes\bm{\Pop}_1}$, followed by the unitary compression stroke on the working substance alone and the second interaction unitary $\Vop_2=e^{i\Hop_2\otimes\bm{\Pop}_1}$ with the same pointer \emph{i.e.} the total unitary evolution of the system and pointer reads:
\begin{align*}
\Vop_2\Uop_1\Vop_1(\hrho_1\otimes\bm{\hrho}_{\sigma_1})\Vop_1^{\dagger}\Uop_1^{\dagger}\Vop_2^{\dagger}.
\end{align*}
It is only after the final interaction that the pointer is measured. Here the first interaction shifts the position of the pointer proportional to the negative of the energy of the system and the second interaction shifts the pointer position proportional to the positive value of the energy of the system, allowing the readout of the difference between the two values \emph{i.e.} the work. In a similar manner, we implement the pointer interactions to measure $q_h$ and $w_3$. After one complete cycle, the non-normalized state of the system conditioned on the pointer measurement outcomes $\vec{x}=(x_1,x_2,x_3)$ reads,
% \begin{align}
% \Phi_{\mathrm{S2},\Vec{x}}(\hrho_1) &= \sum_{\Vec{m},\Vec{m}'} \mathcal{S}^{\Vec{m},\Vec{m}'}(\hrho_1)\rho_{\sigma_1}\left(x_1-w_1^{\vec{m}},x_1-w_1^{\vec{m}'}\right)\nonumber\\
% &\times\rho_{\sigma_2}\left(x_3-w_3^{\vec{m}},x_3-w_3^{\vec{m}'}\right)\nonumber\\
% &\times\rho_{\sigma_3}\left(x_2-q_h^{\vec{m}},x_2-q_h^{\vec{m}'}\right),
% \label{scheme2non-normalisedState}
% \end{align}
\begin{align}
&\Phi_{\mathrm{S2},\Vec{x}}(\hrho_1) = \sum_{\Vec{m},\Vec{m}'} \mathcal{S}^{\Vec{m},\Vec{m}'}(\hrho_1)\Lambda^{\vec{m},\vec{m}'}_{\mathrm{S2}}G_{\sigma_1}\left(x_1-\frac{w_1^{\vec{m}}+w_1^{\vec{m}'}}{2}\right)\nonumber\\
& G_{\sigma_2}\left(x_2-\frac{q_h^{\vec{m}}+q_h^{\vec{m}'}}{2} \right)G_{\sigma_3}\left(x_3-\frac{w_3^{\vec{m}}+w_3^{\vec{m}'}}{2}\right)
\label{scheme2non-normalisedState2}
\end{align}
where $w_1^{\vec{m}}=e_{m_2}^{(2)}-e_{m_1}^{(1)}$, $w_3^{\vec{m}}=e_{m_4}^{(1)}-e_{m_3}^{(2)}$, $q_h^{\vec{m}}=e_{m_3}^{(2)}-e_{m_2}^{(2)}$ (similar definitions for the primed variables), and 
\begin{align*}
\Lambda^{\vec{m},\vec{m}'}_{\mathrm{S2}}=e^{-\frac{1}{8\sigma_1^2}(w_1^{\vec{m}}-w_1^{\vec{m}'})^2-\frac{1}{8\sigma_2^2}(q_h^{\vec{m}}-q_h^{\vec{m}'})^2-\frac{1}{8\sigma_3^2}(w_3^{\vec{m}}-w_3^{\vec{m}'})^2}.
\end{align*}
The normalized state of the system can be obtained by integrating out the pointer variables leading to the cycle CPTP map
\begin{align}
\Phi_{\mathrm{S2}}(\hrho_1) &= \sum_{\Vec{m},\Vec{m}'} \mathcal{S}^{\Vec{m},\Vec{m}'}(\hrho_1)\Lambda^{\vec{m},\vec{m}'}_{\mathrm{S2}}.
\label{scheme2CPTPmap}
\end{align} 
The steady-state for the above CPTP map with the measurement scheme S2 is given by
\begin{align}
\hrho^{\mathrm{ss}}_{\mathrm{S2}}=\Phi_{\mathrm{S2}}(\hrho^{\mathrm{ss}}_{\mathrm{S2}}).
\label{scheme2CPTPmapSS}
\end{align}
With this steady-state, we can write down the characteristic function and work-heat probability distribution respectively as:
%e^{i\left(\frac{k_1}{2}w^{\vec{m},\vec{m}'}+\frac{k_2}{2}q_h^{\vec{m},\vec{m}'}\right)}
\begin{align}
&\chi_{\mathrm{S2}}(k_1,k_2)=\sum_{\Vec{m},\Vec{m}'} \Tr{\mathcal{S}^{\Vec{m},\Vec{m}'}(\hrho^{\mathrm{ss}}_{\mathrm{S2}}) }{s}\Lambda^{\vec{m},\vec{m}'}_{\mathrm{S2}} e^{i\frac{k_1}{2}w^{\vec{m},\vec{m}'}}\nonumber\\
& \times e^{i\frac{k_2}{2}q_h^{\vec{m},\vec{m}'}}e^{-\frac{k_1^2}{2}(\sigma_1^2+\sigma_3^2)-\frac{k_2^2}{2}\sigma_2^2},
\label{scheme2Characteristicfunc}
\end{align}
and
\begin{align}
&p_{\mathrm{S2}}(w,q_h)=\sum_{\Vec{m},\Vec{m}'} \Tr{\mathcal{S}^{\Vec{m},\Vec{m}'}(\hrho^{\mathrm{ss}}_{\mathrm{S2}})}{s}  \Lambda_{\mathrm{S2}}^{\vec{m},\vec{m}'} \nonumber\\
&\times G_{\sqrt{\sigma_1^2+\sigma_2^2}}\left(w-w^{\vec{m},\vec{m}'}/2 \right) G_{\sigma_2}\left(q_h-q_h^{\vec{m},\vec{m}'}/2\right)    \label{scheme2ProbDist}.
\end{align}
% The corresponding joint work-heat probability is,
% \begin{align}
% p_{\mathrm{S2}}(w,q_h) &= \int \prod_{k=1}^{3}dx_k \Tr{\Phi_{\mathrm{S2},\Vec{x}}(\hrho^{\mathrm{ss}}_2)}{s}\delta[w-(x_1+x_3)]\nonumber\\
% &\times \delta[q_h-x_2],
% \label{scheme2ProbDist}
% \end{align} 

% The characteristic function for the joint work and heat distribution evaluated as,

% \begin{align}
% \end{align} 

% From the above expression, we can write down the joint work-heat distribution as,

% \begin{align}
% p_{\mathrm{S2}}(w,q_h)&=\sum_{\Vec{m},\Vec{m}'} \Tr{\mathcal{S}^{\Vec{m},\Vec{m}'}(\hrho^{\mathrm{ss}}_{\mathrm{S2}}) }{s}  \Lambda_{\mathrm{S2}}^{\vec{m},\vec{m}'} \nonumber\\
% &\times G_{\sqrt{\sigma_1^2+\sigma_2^2}}(w-w^{\vec{m},\vec{m}'}/2) G_{\sigma_2}(q_h-q_h^{\vec{m},\vec{m}'}/2),
% \label{scheme2ProbDist2}
% \end{align} 

Comparing the expression of the CPTP map \eqref{scheme2CPTPmap} with \eqref{scheme1CPTPmap}, we see that the key difference between the two measurements schemes is the term which exponentially suppresses the non-diagonal contributions $\vec{m}\neq\vec{m'}$ \cite{Son2021}. In $\mathrm{S1}$, the non-diagonal terms with $\vec{m}\neq\vec{m'}$ are suppressed by the differences of individual energies at the four points of the cycle. While in $\mathrm{S2}$, the non-diagonal terms are exponentially suppressed by the differences in work and heat values for the individual strokes. For scheme $\mathrm{S2}$, we again recover the unmonitored Otto cycle case in the imprecise measurement limit. More specifically, for $\sigma_k\to\infty$, we get $\Lambda_{\mathrm{S2}}^{\Vec{m},\Vec{m'}}\to1$ and the corresponding steady-state tends to that of the unmonitored Otto cycle. In contrast, in the precise pointer limit with $\sigma_k \to 0$ we recover the TPM limit only for a working substance system with non-degenerate work (heat) values \emph{i.e} $\{ w_1^{\Vec{m}}-w_1^{\Vec{m'}},w_3^{\Vec{m}}- w_3^{\Vec{m'}},q_h^{\Vec{m}}-q_h^{\Vec{m'}} \} \neq 0$ for all $\Vec{m}\neq\Vec{m'}$ (except for the trivial cases such as $w_i^{\Vec{m}}=w_i^{\Vec{m'}}=q_h^{\Vec{m}}=q_h^{\Vec{m'}}=0$). Note that this condition of non-degenerate work (heat) values is satisfied for the example of TLS working substance that we will consider in this article. However, this condition will not hold in general for other working substances with multiple energy levels. Finally, let us compare the expression for the joint work-heat distributions of $\mathrm{S2}$ and $\mathrm{S1}$ in Eqs.~\eqref{scheme1ProbDist2} and \eqref{scheme2ProbDist}. We can see that the Gaussian appearing in the sum for scheme $\mathrm{S2}$ has reduced variance compared to $\mathrm{S1}$. This reduction in the variance is due to the reduced number of measurements performed on the pointers in $\mathrm{S2}$ compared to $\mathrm{S1}$. A more detailed comparison for the specific TLS working substance will be made later. 

\subsection{\label{subsec:Scheme 3} Measurement Scheme 3}
In the third scheme $\mathrm{S3}$, we measure the total work $w=w_1+w_3$ and heat exchanged during the hot isochore $q_h$. For this we only need two pointers and two measurements. For the measurement of total work $w$, we need four interactions between system and pointer at four points of the Otto cycle and one measurement at the end of the stroke ($3\to4$) as shown schematically in Fig.~(\ref{fig:MeasurementSchemeSchematic}c). In addition, for the measurement of heat exchange $q_h$, we need two interactions between the system and pointer before and after the hot isochore ($2\to3$) followed by the measurement at point $3$. After one complete cycle, the conditional non-normalized state of the system conditioned on the two measurement outcomes $\vec{x}=(x_1,x_2)$ reads
% ,
% \begin{align}
% \Phi_{\mathrm{S3},\vec{x}}(\hrho_1) &= \sum_{\Vec{m},\Vec{m}'} \mathcal{S}^{\Vec{m},\Vec{m}'}(\hrho_1)\rho_{\sigma_1}\left(x_1-w^{\vec{m}},x_1-w^{\vec{m}'}\right)\nonumber\\
% &\times\rho_{\sigma_2}\left(x_2-q_h^{\vec{m}},x_2-q_h^{\vec{m}'}\right).
% \label{scheme3non-normalisedState}
% \end{align}
% Which can be simplified to,
\begin{align}
&\Phi_{\mathrm{S3},\vec{x}}(\hrho_1)  = \sum_{\Vec{m},\Vec{m}'} \mathcal{S}^{\Vec{m},\Vec{m}'}(\hrho_1)\Lambda^{\vec{m},\vec{m}'}_{\mathrm{S3}}\nonumber\\
&\times G_{\sigma_1}(x_1-(w^{\vec{m}}+w^{\vec{m}'})/2) G_{\sigma_2}(x_2-(q_h^{\vec{m}}+q_h^{\vec{m}'})/2),
\label{scheme3non-normalisedState2}
\end{align}
with $\Lambda^{\vec{m},\vec{m}'}_{\mathrm{S3}}=e^{-\frac{1}{8\sigma_1^2}(w^{\vec{m}}-w^{\vec{m}'})^2-\frac{1}{8\sigma_2^2}(q_h^{\vec{m}}-q_h^{\vec{m}'})^2}$. The normalized state of the system can be obtained by integrating out the pointer variables leading to the CPTP map for scheme S3
\begin{align}
\Phi_{\mathrm{S3}}(\hrho_1) &= \sum_{\Vec{m},\Vec{m}'} \mathcal{S}^{\Vec{m},\Vec{m}'}(\hrho_1)\Lambda^{\vec{m},\vec{m}'}_{\mathrm{S3}},
\label{scheme3CPTPmap}
\end{align} 
and the associated steady state
\begin{align}
\hrho^{\mathrm{ss}}_{\mathrm{S3}}=\Phi_{\mathrm{S3}}(\hrho^{\mathrm{ss}}_{\mathrm{S3}}).
\label{scheme3CPTPmapSS}
\end{align}
The corresponding characteristic function and probability distributions for the work and heat probability read
\begin{align}
&\chi_{\mathrm{S3}}(k_1,k_2)=\sum_{\Vec{m},\Vec{m}'}
\Tr{\mathcal{S}^{\Vec{m},\Vec{m}'}(\hrho^{\mathrm{ss}}_{\mathrm{S3}})}{s}\Lambda^{\vec{m},\vec{m}'}_{\mathrm{S3}} \nonumber\\
&\times e^{i\frac{k_1}{2}w^{\vec{m},\vec{m}'}+i\frac{k_2}{2}q_h^{\vec{m},\vec{m}'}} e^{-\frac{k_1^2}{2}\sigma_1^2-\frac{k_2^2}{2}\sigma_2^2} ,
\label{scheme3Characteristicfunc}
\end{align} 
and
\begin{align}
&p_{\mathrm{S3}}(w,q_h)=\sum_{\Vec{m},\Vec{m}'} \Tr{\mathcal{S}^{\Vec{m},\Vec{m}'}(\hrho^{\mathrm{ss}}_{\mathrm{S3}}) }{s}  \Lambda_{\mathrm{S3}}^{\vec{m},\vec{m}'} \nonumber\\
&\times G_{\sigma_1}(w-w^{\vec{m},\vec{m}'}/2) G_{\sigma_2}(q_h-q_h^{\vec{m},\vec{m}'}/2).
\label{scheme3ProbDist2}
\end{align} 
From the expressions of the CPTP map and work and heat probability distribution for the scheme $\mathrm{S3}$, we see that here the non-diagonal terms $\vec{m}\neq\Vec{m'}$ are suppressed by the difference of total work values i.e $w^{\Vec{m}}$ and $w^{\Vec{m'}}$ and heat values i.e $q_h^{\Vec{m}}$ and $q_h^{\Vec{m'}}$. Similarly to the previous discussion, we recover the unmonitored Otto cycle case in the imprecise measurement limit ($\sigma_k\to\infty$). While on the other hand the non-diagonal terms  $\vec{m}\neq\Vec{m'}$ with the same work output i.e $w^{\Vec{m}}-w^{\Vec{m'}}=0$ will not be suppressed in the limit $\sigma_k\to0$. Note that, unlike the S2 scheme where this was not possible with a TLS system, irrespective of the spectrum of the working substance  there will be multiple processes where the work in the compression will equal the negative of the work in the expansion. For instance this is the case where total work is zero $w^{\Vec{m}}=w^{\Vec{m'}}=0$ but $\Vec{m} \neq \Vec{m'}$. As a result, the steady-state and the joint work-heat distribution for scheme $\mathrm{S3}$ differ from the TPM values even in the limit $\sigma_k \to 0$ in general. We will demonstrate with explicit example working substance in the next section.% More specifically for $\sigma_k>>\mathrm{max}_{\Vec{m},\Vec{m'}}\{|w^{\Vec{m}}-w^{\Vec{m'}}|,|q_h^{\Vec{m}}-q_h^{\Vec{m'}}|\}$, we get $\Lambda_{\mathrm{S3}}^{\Vec{m},\Vec{m'}}\to1$ and the corresponding steady-state tend to that of the unmonitored Otto cycle.
\section{\label{sec:Results}Results}
In the previous section we have discussed how the three different measurement schemes (S1,S2,S3) lead to different probability distributions for work and heat for a generic working substance. While this approach yields some insights, it is not possible to obtain analytical expressions for the steady state density matrix of the cycle CPTP maps defined via Eqs.~\eqref{scheme1CPTPmapSS},~\eqref{scheme2CPTPmapSS},~\eqref{scheme3CPTPmapSS} for generic working substances. In this section by choosing a specific working substance, namely a two-level system, we will determine the steady states for the different schemes (numerically) and thus quantitatively demonstrate the impact of the measurement scheme choice on the average work output of the quantum Otto heat engine. We begin by describing the TLS and the protocol we use.
\subsection{\label{subsec:System}Two-Level Working Substance}
We consider a two-level system with time-dependent Hamiltonian $\Hop(t)$ which take the following general form at the beginning of the compression and expansion stroke respectively:
\begin{align}
  \Hop_1 &= e_{1}^{(1)} \ket{e_{1}^{(1)}}\bra{e_{1}^{(1)}} + e_{2}^{(1)} \ket{e_{2}^{(1)}}\bra{e_{2}^{(1)}} \label{TLSHamiltonian1}\\
\Hop_2 &= e_{1}^{(2)} \ket{e_{1}^{(2)}}\bra{e_{1}^{(2)}} + e_{2}^{(2)} \ket{e_{2}^{(2)}}\bra{e_{2}^{(2)}}\label{TLSHamiltonian2}
\end{align}
Recall that during the compression stroke the hamiltonian is changed over the time $\tau_u$ from $\Hop_1$ to $\Hop_2$ and the expansion stroke is generated by the precise time-reverse of the compression stroke. While we will present one realization of these strokes for a particular choice of $\Hop_1$ and $\Hop_2$ in Sec.~\eqref{subsec:Numerics}, we can in general write the unitary operators corresponding to the compression ($\Uop_1$) and expansion ($\Uop_2 = \hat{\Theta} \Uop_1^{\dagger} \hat{\Theta}^{\dagger}$, with $\hat{\Theta}$ the time-reversal operator) strokes as \cite{Son2021}
\begin{align}
\Uop_1 &= \sqrt{1-r}\left(\ket{e_2^{(2)}}\bra{e_2^{(1)}}+\ket{e_1^{(2)}}\bra{e_1^{(1)}}\right)\nonumber\\
&-\sqrt{r}\left(e^{i\phi}\ket{e_2^{(2)}}\bra{e_1^{(1)}}-e^{-i\phi}\ket{e_1^{(2)}}\bra{e_2^{(1)}}\right)\label{TLSUnitary1}\\
\Uop_2 &= \sqrt{1-r}\left(\ket{e_2^{(1)}}\bra{e_2^{(2)}}+\ket{e_1^{(1)}}\bra{e_1^{(2)}}\right)\nonumber\\
&+\sqrt{r}\left(e^{i\phi}\ket{e_2^{(1)}}\bra{e_1^{(2)}}-e^{-i\phi}\ket{e_1^{(1)}}\bra{e_2^{(2)}}\right)
\label{TLSUnitary2}.
\end{align} 
Here the transition probability $r\in[0,1]$ and the associated transition amplitude's phase $\phi\in[0,2\pi]$ are determined by the details of the protocol implementing the strokes. Note that $r \to 0$ corresponds to the quasistatic limit with no transitions between the ground and excited states. 

The dissipative dynamics is modeled in the limit of weak system-bath coupling by a Gorini-Kossakowski-Sudarshan-Lindblad (GKSL) Markovian master equation,
\begin{align}
\frac{d\hrho}{dt}&=-i\commu{\Hop_i}{\hrho} + \gamma_i(n^{(i)}_{b}+1)\left(\sigma_-^{(i)}\hrho\sigma_{+}^{(i)} - \frac{1}{2}\{\sigma_{+}^{(i)}\sigma_{-}^{(i)},\hrho \}\right)\nonumber\\
&+\gamma_in^{(i)}_{b}\left(\sigma_{+}^{(i)}\hrho\sigma_{-}^{(i)} - \frac{1}{2}\{\sigma_{-}^{(i)}\sigma_{+}^{(i)},\hrho \}\right),
\label{TLSMaster}
\end{align} 
with $\sigma_{+}^{(i)}=\sigma_{-}^{(i){\dagger}} = \ket{e_2^{(i)}}\bra{e_1^{(i)}}$ between eigenstates of the Hamiltonian $\Hop_i$ ($i=1,2$). Note that $\{ \cdot,\cdot\}$ denotes the anti-commutator. The thermal occupation numbers for the hot and the cold bath read $n_b^{(2)}=\frac{1}{e^{\beta_h(e_2^{(2)} - e_1^{(2)})}+1}$ and $n_b^{(1)}=\frac{1}{e^{\beta_c(e_2^{(1)} - e_1^{(1)})}+1}$. Integrating the master equation \eqref{TLSMaster} gives a qunatum map $\Phi_{\beta_{c}}(.)$ and $\Phi_{\beta_{h}}(.)$ corresponding to two temperatures $\beta_c$ and $\beta_h$. This CPTP map brings any initial system density matrix to the respective Gibbs thermal state in the limit of long thermalization strokes $\tau_b \rightarrow \infty$.
%%%%%%%%%%
\begin{figure*}
	\centering
	\subfloat{\begin{overpic}[width=0.33 \linewidth]{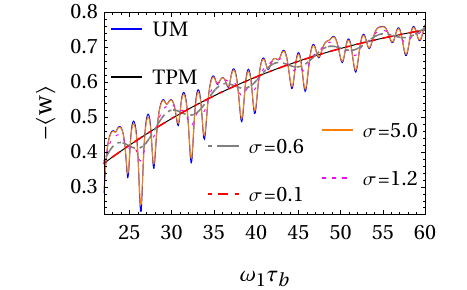}
	\put(45,53){\textbf{(a)}}
	\end{overpic}
	}
	\subfloat{\begin{overpic}[width=0.33 \linewidth]{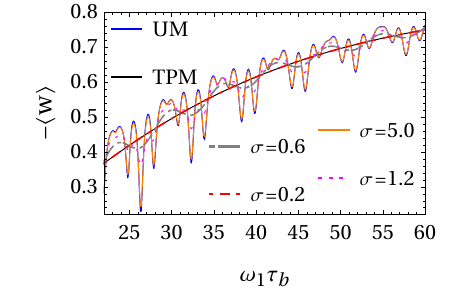}
	\put(45,53){\textbf{(b)}}
	\end{overpic}
	}
	\subfloat{\begin{overpic}[width=0.33 \linewidth]{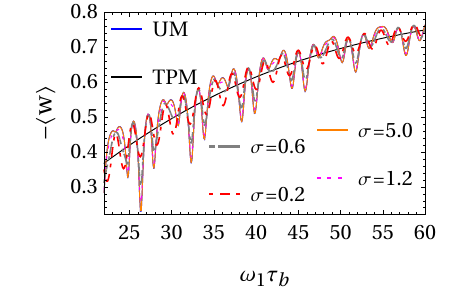}
	\put(45,53){\textbf{(c)}}
	\end{overpic}
	}\\ 
    \caption{(Color Online) Comparison plot of average work for $\mathrm{S1}$, $\mathrm{S2}$ and $\mathrm{S3}$ along with corresponding values for UM and TPM as a function of thermalization time $\omega_1\tau_b$ (a), (b) and (c) for fixed value of the unitary stroke time $\omega_1\tau_u=3.5$ for different values of $\sigma$. Other parameter values are $\omega_1=1.0$, $\omega_2=3.2$, $\beta_c=3.0$, $\beta_h=0.2$, $\gamma_h=0.05$ and $\gamma_c=0.05$. 
  }
    \label{fig:ComparisonOfwavg}
\end{figure*}
%%%%%%%%%%%%%%%%%%%%%%%
\subsection{\label{subsec:Perfectly Thermalising Cooling Isochore}Perfectly Thermalising Cooling Isochore}

With the choice of a TLS as a working substance hamiltonian and the compression and expansion work protocols given by Eqs.~\eqref{TLSUnitary1} and \eqref{TLSUnitary2} and the thermal strokes modeled by \eqnref{TLSMaster}, Eqs.~\eqref{scheme1CPTPmapSS},~\eqref{scheme2CPTPmapSS},~\eqref{scheme3CPTPmapSS} reduce to fixed point equations for the matrix elements of the steady-state density matrix that parametrically depend on the system parameters and the pointer widths $\sigma_i$. Even for the simple TLS working substance case these equations are cumbersome and an exact analytical solution is intractable. Before we present results from numerical solutions, in order to obtain some analytical insights we consider now a simpler scenario of the Otto cycle with the cooling isochore $4\to1$ to be perfectly thermalizing \cite{Camati2019}. Without loss of generality we take $e_2^{1}=-e_1^{1}=\omega_1/2$ and $e_2^{2}=-e_1^{2}=\omega_2/2$. Also from this point on, for simplicity we take the pointer widths in the different strokes to be uniform \emph{i.e.} $\sigma_i = \sigma$ in all the schemes.

For a perfectly thermalizing cooling iscochore, the system reaches a Gibbs state at the end of the cooling stroke. In this case the Gibbs state $\hrho_{\beta_c}=\frac{e^{-\beta_c\Hop_2}}{Z_c}$ corresponding to the inverse temperature $\beta_c$ becomes the steady state of the Otto cycle. Note that the hot isochore is still of finite duration and hence not perfectly thermalizing. Due to the imperfect thermalization during the heating isochore, the coherence generated in the non-quasistatic compression stroke does not decay to zero. In this case, we can write down the analytical expressions for the characteristic functions corresponding to the different schemes. Since the exact expressions for the characteristic functions are cumbersome and do not provide insight, we directly analyze the average work output (and its fluctuations) in the different measurement schemes next.

The average TPM and unmonitored work, $\langle w\rangle_{\mathrm{TPM}}$ and $\langle w\rangle_{\mathrm{UM}}$, respectively are given by
\begin{align}
&\langle w\rangle_{\mathrm{TPM}} = (1-e^{-\gamma_h\tau_b})\left[(1-2r)\omega_1-\omega_2\right]n_b^{(h)}\nonumber\\
&+\left[\omega_2(1-2r)-\omega_1\right]n_b^{(c)} +r(\omega_2+\omega_1)\nonumber\\
&-e^{-\gamma_h\tau_b}\left[r+(1-2r) n_b^{(c)}\right]\left[\omega_2-(1-2r)\omega_1\right],
\label{wavgTPM}\\
&\langle w\rangle_{\mathrm{UM}} = \langle w\rangle_{\mathrm{TPM}}\nonumber\\
& - 2e^{-\frac{1}{2}\gamma_h\tau_b}r(1-r)\omega_1\cos\left(2q+\omega_2\tau_b\right)\tanh\left(\frac{\beta_c\omega_1}{2}\right).
\label{wavgPerfectCooling}
\end{align}
%%%%%%%%%%%%%%%%%%%%%%  
\begin{figure*}
	\centering
	\subfloat{\begin{overpic}[width=0.33 \linewidth]{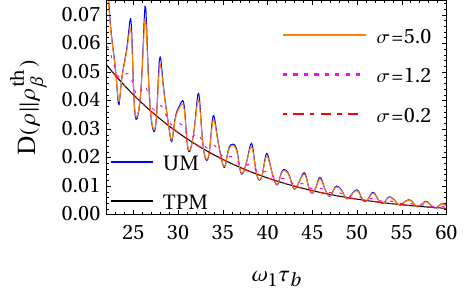}
	\put(40,55){\textbf{(a)}}
	\end{overpic}
	}
	\subfloat{\begin{overpic}[width=0.33 \linewidth]{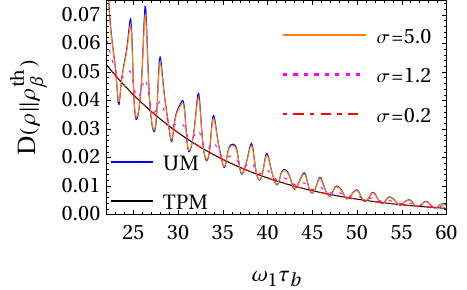}
	\put(40,55){\textbf{(b)}}
	\end{overpic}
	}
	\subfloat{\begin{overpic}[width=0.33 \linewidth]{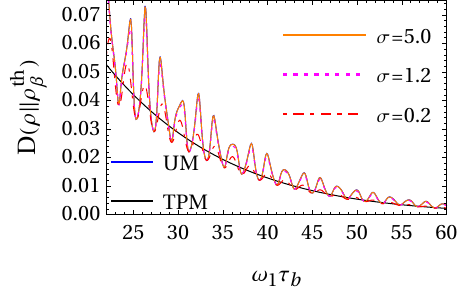}
	\put(40,55){\textbf{(c)}}
	\end{overpic}
	}\\  
    \caption{(Color Online) 
Comparison plot of Kullback-Leibler divergence between Gibbs thermal state $\hrho^{\mathrm{th}}_{\beta_c}$ and steady-state corresponding to $\mathrm{S1}$, $\mathrm{S2}$ and $\mathrm{S3}$ along with corresponding values for UM and TPM as a function of thermalization time $\omega_1\tau_b$ (a), (b) and (c) for different values of $\sigma$. Other parameter values are same as in \figref{fig:ComparisonOfwavg}.  }
     \label{fig:ComparisonOfDivergences}
\end{figure*}
%%%%%%%%%%%%%%%%%%%%%%%%%%%%%
From the above equations, we see that while $\langle w\rangle_{\mathrm{UM}}$ oscillates as a function of the hot isochore duration $\tau_b$, $\langle w\rangle_{\mathrm{TPM}}$ is monotonic. As discussed in \cite{Camati2019}, it is evident from Eq.~\eqref{wavgPerfectCooling} that this oscillation arises from a combination of coherence generation in the work strokes that requires $r \neq 1$ (non quasi-static) and imperfect thermalization engendered by a finite duration $\gamma \tau_b$ of the hot isochore. Moreover, even though these features are necessary for the oscillation they are not sufficient as we can see that the projective energy measurements kill the coherence in the TPM case. 
%As discussed in \cite{Camati2019}, this oscillation is due to imperfect thermalization during the hot isochore as opposed to the case of TPM which shows no oscillation since the projective energy at each point of the Otto cycle kills the energy coherence. Note that the oscillations in $\avg{w}_{\mathrm{UM}}$ damps out as we make the thermalization time $\tau_b$ large. Also, we need $r\neq1$, i.e. non-quasistatic work strokes leading to coherence generation, for the non-zero oscillation. Thus the work strokes have to be non-quasistatic such that there is coherence generation in the instantaneous energy basis during unitary work strokes. Moreover, note that the amplitude of oscillations in the unmonitored work is significant only in the low temperature $\beta_c \omega_1 \gg 1$ regime
One of the consequences of this is that the unmonitored work can be lower or higher than the corresponding TPM value depending on the duration of the heating isochore. More specifically from \eqref{wavgPerfectCooling} we see that $\avg{w}_{\mathrm{UM}}>\avg{w}_{\mathrm{TPM}}$ for $\pi\leq2q+\omega_2\tau_b\leq2\pi$. The expression of average work for the schemes $\mathrm{S1}$, $\mathrm{S2}$ and $\mathrm{S3}$ can be written down as,
\begin{align}
\langle w \rangle_{\mathrm{S1}}=\langle w \rangle_{\mathrm{S2}} &= e^{-\frac{\omega_2^2}{4\sigma^2}}\langle w \rangle_{\mathrm{UM}} \nonumber \\
& + \left(1-e^{-\frac{\omega_2^2}{4\sigma^2}}\right)\langle w \rangle_{\mathrm{TPM}} \label{wavgS1S2} \\
\langle w \rangle_{\mathrm{S3}}&=\langle w \rangle_{\mathrm{UM}}.
\label{wavgS3}
\end{align} 
From Eq.~\eqref{wavgS1S2} we see that, $\langle w \rangle_{\mathrm{S1}}$ and $\langle w \rangle_{\mathrm{S2}}$ interpolates between two limiting values $\langle w \rangle_{\mathrm{UM}}$ and $\langle w \rangle_{\mathrm{TPM}}$ as we tune pointer width $\sigma$. More specifically for $\sigma \gg \omega_2$ we recover the unmonitored limit, while for $\sigma \ll \omega_2$ we recover the TPM limit. On the other hand, $\langle w \rangle_{\mathrm{S3}}$ is independent of the measurement and remains equal to $\langle w \rangle_{\mathrm{UM}}$. This can be reasoned as follows. In $\mathrm{S3}$ the measurement of total work is performed only after the completion of the expansion work stroke $3\to4$. Although there will be a measurement back-action due to the final measurement, the final cooling stroke maps the resultant state back to the Gibbs state. Also the measurement of heat after the completion of stroke $2\to3$ does not affect the measurement of work in this case. We note that this is because of the special choice of the dissipative dynamics that uncouples the evolution of populations and coherences \cite{Son2021}.

Coming to the fluctuations of work in the different schemes, we find that the second cumulant of work take the following form (compared to the TPM case)
%\begin{align}
%&\langle w^2 \rangle_{\mathrm{S1}}=\langle w^2 \rangle_{\mathrm{S2}} = \langle w^2 \rangle_{\mathrm{TPM}} + 4\sigma^2 \nonumber  \\ 
%&- 2\omega_1^2e^{-\frac{\gamma_h\tau_b}{2}}e^{-\frac{\omega_2^2}{4\sigma^2}}r(1-r)\cos\left(2\phi+\omega_2\tau_b\right) \\
%&\langle w^2 \rangle_{\mathrm{S3}} = \langle w^2 \rangle_{\mathrm{TPM}}+\sigma^2 \nonumber\\
%&-2\omega_1^2e^{-\frac{\gamma_h\tau_b}{2}}r(1-r)\cos\left(2\phi+\omega_2\tau_b\right) .
%\label{secondmoments}
%\end{align} 
\begin{align}
&\avg{w^2}_c^{\mathrm{S1}}=\avg{w^2}_c^{\mathrm{TPM}} +  4\sigma^2 -2e^{-\frac{\omega_2^2}{4\sigma^2}-\frac{\gamma_h\tau_b}{2}}r(1-r) \nonumber\\
&\times\left[\omega_1^2-2\omega_1\avg{w}_{\mathrm{TPM}}\tanh\left(\frac{\beta_c\omega_1}{2}\right)\right]\cos(2\phi+\omega_2\tau_b)\nonumber\\
&-4e^{-\frac{\omega_2^2}{2\sigma^2}-\gamma_h\tau_b}\tanh^2\left(\frac{\beta_c\omega_1}{2}\right)\omega_1^2r^2(1-r)^2\cos^2(2\phi+\omega_2\tau_b)\\
&\avg{w^2}_c^{\mathrm{S2}}=\avg{w^2}_c^{\mathrm{TPM}} + 2\sigma^2 -2e^{-\frac{\omega_2^2}{4\sigma^2}-\frac{\gamma_h\tau_b}{2}}\nonumber\\
&\times\left[\omega_1^2-2\omega_1\avg{w}_{\mathrm{TPM}}\tanh\left(\frac{\beta_c\omega_1}{2}\right)\right]r(1-r)\cos(2\phi+\omega_2\tau_b) \nonumber\\
&- 4e^{-\frac{\omega_2^2}{2\sigma^2}-\gamma_h\tau_b}\tanh^2\left(\frac{\beta_c\omega_1}{2}\right)\omega_1^2r^2(1-r)^2\cos^2(2\phi+\omega_2\tau_b)\\
&\avg{w^2}_c^{\mathrm{S3}}=\avg{w^2}_c^{\mathrm{TPM}}+\sigma^2-2e^{-\frac{\gamma_h\tau_b}{2}}\nonumber\\
&\times\left[\omega_1^2-2\omega_1\avg{w}_{\mathrm{TPM}}\tanh\left(\frac{\beta_c\omega_1}{2}\right)\right]r(1-r)\cos(2\phi+\omega_2\tau_b) \nonumber\\
&- 4e^{-\gamma_h\tau_b}\tanh^2\left(\frac{\beta_c\omega_1}{2}\right)\omega_1^2r^2(1-r)^2\cos^2(2\phi+\omega_2\tau_b)
\label{secondmoments}
\end{align} 

We can immediately see from the expressions that the weak pointer-based measurements contribute additional noise compared to the TPM. For large pointer noise $\sigma$, the variance square of all three schemes is dominated by the terms proportional to $\sigma^2$. In fact, the square of variance is largest for $\mathrm{S1}$ and smallest for $\mathrm{S3}$. As discussed before for the general measurement scheme setup, this additional noise is coming from the number of pointer measurements done in each scheme. In the precise measurement limit $\sigma\to0$, the variance for $\mathrm{S1}$ and $\mathrm{S2}$ approaches the TPM value while this is not the case for $\mathrm{S3}$. More precisely the variance for $\mathrm{S3}$ in the limit $\sigma\to0$ deviates from the TPM values and it can be larger or smaller depending on the duration of the hot isochore $\tau_b$ and unitary stroke parameters $r$ and $\phi$. Also, for finite but small values of $\sigma$, we have found by sweeping over different parameter regimes and plotting in the different parameter regimes that the variance of $\mathrm{S1}$ and $\mathrm{S2}$ are always lower bounded by TPM value whereas the variance for $\mathrm{S3}$ can take values smaller than TPM.

\subsection{\label{subsec:Numerics}Numerics}
%%%%%
\begin{figure*}
	\centering
	\subfloat{\begin{overpic}[width=0.33 \linewidth]{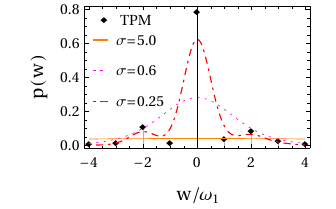}
	\put(78,50){\textbf{(a)}}
	\end{overpic}
	}
	\subfloat{\begin{overpic}[width=0.33 \linewidth]{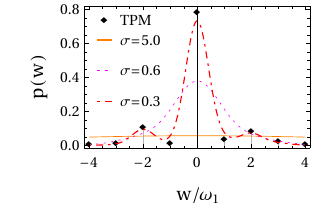}
	\put(78,50){\textbf{(b)}}
	\end{overpic}
	}
	\subfloat{\begin{overpic}[width=0.33 \linewidth]{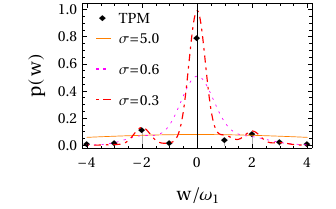}
	\put(78,50){\textbf{(c)}}
	\end{overpic}
	}\\ 
    \caption{(Color Online) Comparison plot of work distribution for different values of $\sigma$ for $\mathrm{S1}$ (a), $\mathrm{S2}$ (b) and $\mathrm{S3}$ (c) along with the TPM limit. Other parameter values are $\omega_1=1.0$, $\omega_2=3.0$, $\beta_c=3.0$, $\beta_h=0.2$, $\gamma_h=0.05$, $\gamma_c=0.05$, $\tau_u=3.5$ and $\tau_b=22.0$. }
        \label{fig:ProbDist1}
\end{figure*}%%%%%%
%%%%%%%%%%%%%%
\begin{figure}
	\centering
	\subfloat{\begin{overpic}[width=0.74 \linewidth]{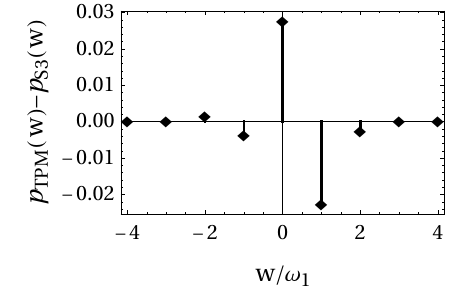}
	\put(28,50){\textbf{(a)}}
	\end{overpic}
	}\\
	\subfloat{\begin{overpic}[width=0.74 \linewidth]{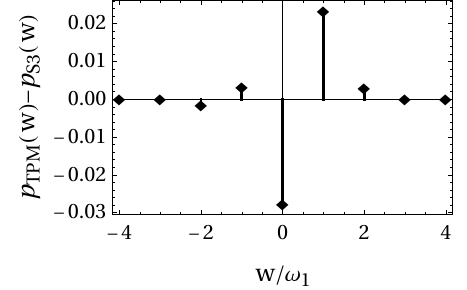}
	\put(28,50){\textbf{(b)}}
	\end{overpic}
	}
	\\ 
    \caption{(Color Online) Work distribution of $\mathrm{S3}$ compared to TPM (for $\sigma\to0$) for $\omega_1\tau_u=3.5$ (a) and $\omega_1\tau_b=25.1$ (b). Here (a) shows the case in which $-\avg{w}_{\mathrm{TPM}}>-\avg{w}_{\mathrm{S3}}$ while (b) shows the case for which $-\avg{w}_{\mathrm{TPM}}<-\avg{w}_{\mathrm{S3}}$. Other parameter values are same as \figref{fig:ProbDist1}}.
     \label{fig:ProbDist2}
\end{figure}

 \begin{figure*}
	\centering
	\subfloat{\begin{overpic}[width=0.33 \linewidth]{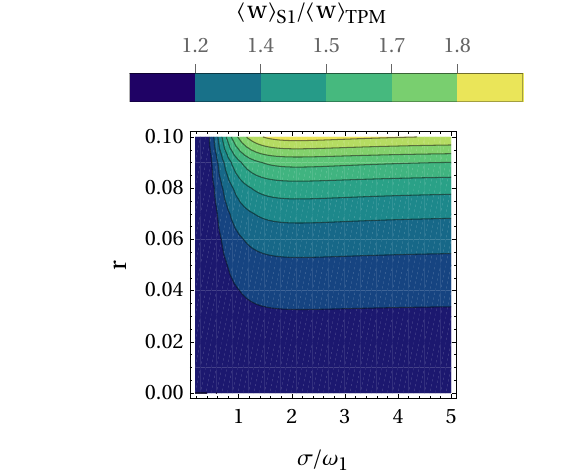}
	\put(70,53){\textbf{(a)}}
	\end{overpic}
	}
	\subfloat{\begin{overpic}[width=0.33 \linewidth]{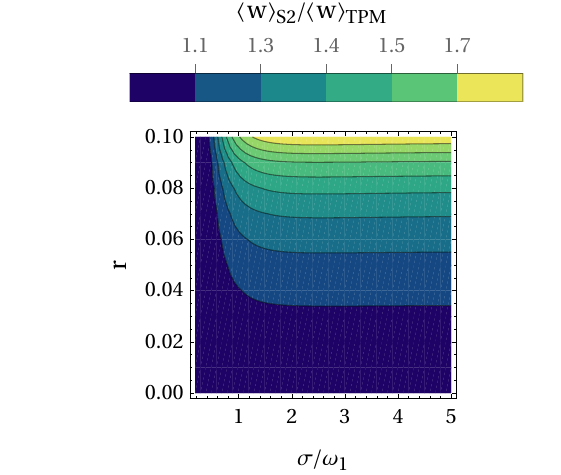}
	\put(70,53){\textbf{(b)}}
	\end{overpic}
	}
	\subfloat{\begin{overpic}[width=0.33 \linewidth]{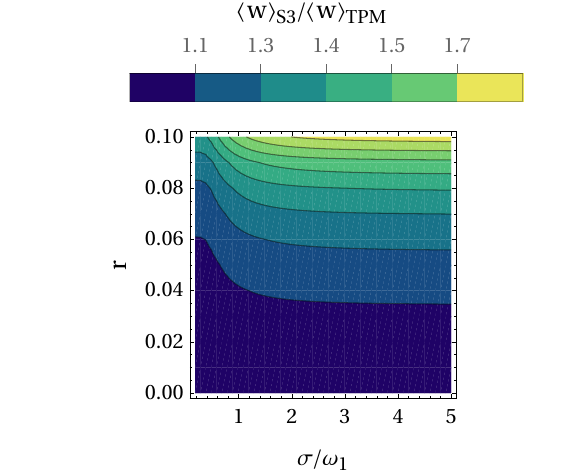}
	\put(70,53){\textbf{(c)}}
	\end{overpic}
	}\\ 
    \caption{(Color Online) Comparison plot of average work (normalized by the TPM average work)  for different values of pointer width $\sigma$ and unitary parameter $r$ for $\mathrm{S1}$ (a), $\mathrm{S2}$ (b) and $\mathrm{S3}$ (c)  for the general two parameters unitary protocol \eqref{TLSUnitary1} with Hamilonian $\Hop_1=\frac{\omega_1}{2}\sigma_z+\frac{\epsilon}{2}\sigma_x$ and $\Hop_2=\frac{\omega_2}{2}\sigma_z+\frac{\epsilon}{2}\sigma_x$. Parameters values are $\omega_1=1.0$, $\omega_2=3.2$, $\epsilon=1.0$, $\phi=\frac{\pi}{5}$, $\beta_c=3.0$, $\beta_h=0.2$, $\gamma_h=\gamma_c=0.05$, and $\tau_b=30.0$. }
        \label{fig:avgworkGenUnitary}
\end{figure*}%%%%%%
%%%%%%%%%%%%%%%%%%
% \begin{align}
% \Hop(t)=\frac{\lambda(t)}{2}\sigma_z+\frac{\epsilon(t)}{2}\sigma_x.
% \label{TLSHamiltonian}
% \end{align} 
% While we expect (and have checked \pvcheck{Attention Rahul}) the results we present are qualitatively valid for arbitrary functions $\lambda(t),\epsilon(t)$, for the sake of simplicity following \cite{Camati2019} (\pvcheck{attention Rahul}), the time-dependent control parameters are chosen such that for the compression work stroke $1\rightarrow 2$ the Hamiltonian is varied from $\Hop(t_1)=\Hop_c=\frac{\omega_c}{2}\sigma_x$ to $\Hop(t_2)=\Hop_h=\frac{\omega_h}{2}\sigma_z$. And for the expansion stroke $3\rightarrow 4$ we have, $\Hop(t_3)=\Hop_h=\frac{\omega_h}{2}\sigma_z$ to $\Hop(t_4)=\Hop_c=\frac{\omega_c}{2}\sigma_x$ ($\omega_h>\omega_c$). Given the specific time-dependent function satisfying the above boundary conditions for the compression stroke, one can write down the general unitary time evolution operators for the compression and expansion work strokes as,
In this section, we consider the Otto cycle with a TLS working system and all four strokes of finite time. For the unitary work strokes we choose a time-dependent protocol as used in \cite{Camati2019}, namely
\begin{align}
\Hop(t)=\frac{\lambda(t)}{2}\left[ \sin \left(\frac{\pi t}{2\tau_u}{}\right)\sigma_z+\cos \left (\frac{\pi t}{2\tau_u} \right) \sigma_x \right],
\label{TLSHamiltoniancomp}
\end{align}
with $\lambda(t) = \omega_1 (1-t/\tau_u) + \omega_2 t/\tau_u$ during compression and its time reverse 
\begin{align}
\Hop(t)=\frac{\lambda(\tau_u-t)}{2}\left[ \cos \left(\frac{\pi t}{2\tau_u}{}\right)\sigma_z+\sin \left (\frac{\pi t}{2\tau_u} \right) \sigma_x \right],
\label{TLSHamiltonianexp}
\end{align}
during the expansion stroke. Note that in both strokes $t \in [0,\tau_u]$. This means that the Hamiltonian is varied (taking $t_1 = 0$) from $\Hop(t_1)=\Hop_1 = \frac{\omega_1}{2}\sigma_x$ to $\Hop(t_2)=\Hop_2=\frac{\omega_2}{2}\sigma_z$. We consider the Otto cycle operating as a heat engine with positive average work extraction (i.e. $-\avg{w}>0$) and positive heat absorption (i.e. $\avg{q_h}>0$). The average values are normalized by corresponding quasistatic limiting values for the cycle (i.e. $\tau_u\to\infty$ and $\tau_b \to \infty$) throughout the sub-section. The key numerical task after specifying the form of the hamiltonian and the corresponding thermalization quantum maps via the GKLS master equations [see \eqref{TLSMaster}] is to construct the cycle maps and numerically solve Eqs.~\eqref{scheme1CPTPmapSS},~\eqref{scheme2CPTPmapSS},~\eqref{scheme3CPTPmapSS} for the corresponding steady states in the different measurement schemes. We have done this for some exemplary parameter choices for the system and note that the results remain qualitatively the same for any parameter choice.

%%%%%%%%%%%%%%%%%%%%%
 \begin{figure*}
	\centering
	\subfloat{\begin{overpic}[width=0.33 \linewidth]{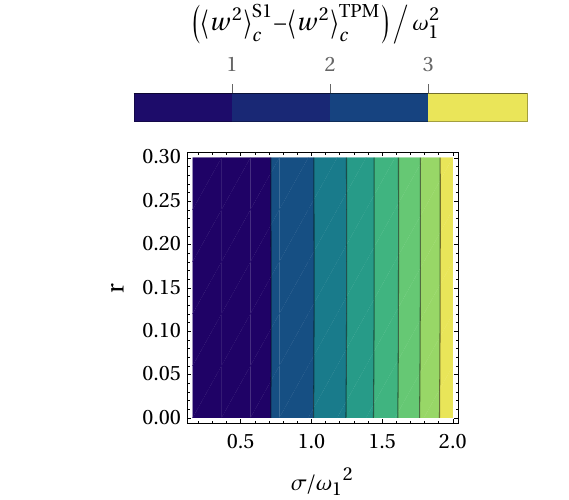}
	\put(70,53){\textbf{(a)}}
	\end{overpic}
	}
	\subfloat{\begin{overpic}[width=0.33 \linewidth]{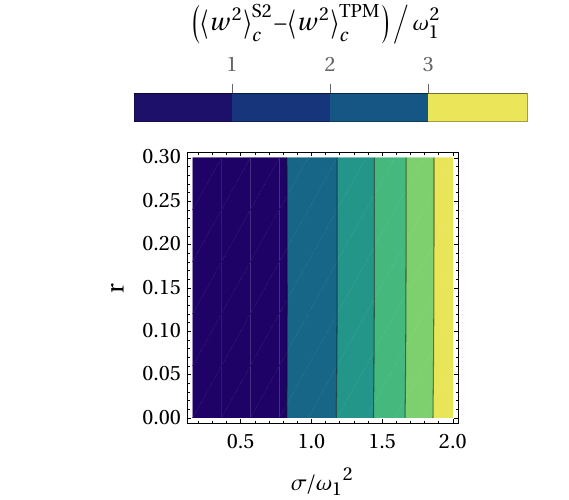}
	\put(70,53){\textbf{(b)}}
	\end{overpic}
	}
	\subfloat{\begin{overpic}[width=0.33 \linewidth]{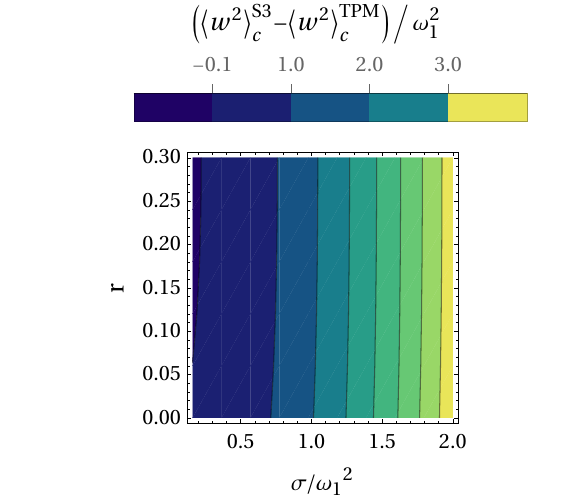}
	\put(70,53){\textbf{(c)}}
	\end{overpic}
	}\\ 
    \caption{(Color Online) Comparison plot of the difference of variance of work for different values of pointer width $\sigma$ and unitary parameter $r$ for $\mathrm{S1}$ (a), $\mathrm{S2}$ (b) and $\mathrm{S3}$ (c)  for the general two parameters unitary protocol \figref{fig:avgworkGenUnitary}. }
        \label{fig:workfluctuationsGenUnitary}
\end{figure*}%%%%%%

Fig.~\eqref{fig:ComparisonOfwavg} shows the plot of average work output as a function of thermalization time $\tau_b$ (such that there is generation of coherence) for measurement schemes $\mathrm{S1}$, $\mathrm{S2}$ and $\mathrm{S3}$ along with UM and TPM values. The average work values for the three measurement schemes are plotted for few different values of pointer width $\sigma$. In Fig.~\eqref{fig:ComparisonOfwavg} (a-c) we see that, the average work output for the UM Otto $\avg{w}_{\mathrm{UM}}$ shows damped oscillation as a function of thermalization time $\tau_b$. While on the other hand the average work output for TPM limit $\avg{w}_{\mathrm{TPM}}$ does not show any oscillations. This behaviour is in agreement (qualitatively) with our discussion in the previous sub-section and as before the reason for the oscillations is precisely the presence of energy coherence before the onset of the isochoric strokes which is absent in the TPM scheme due to the projective energy measurements. Note that while with perfectly thermalizing cooling stroke, the oscillations occur at a single frequency $\omega_h$, here we see a beating like pattern with multiple frequencies. This is because here we get is a combined effect of the damped oscillations coming from imperfect thermalization during both the finite-duration isochoric strokes. 

Focusing on the work output for the measurement schemes $\mathrm{S1}$, $\mathrm{S2}$ presented in Fig.~\eqref{fig:ComparisonOfwavg} (a-b), we see that the average work output interpolates between the TPM and UM values as $\sigma$ is tuned from small to large values. In fact, we see that once the pointer uncertainity exceeds the largest energy gap during the engine cycle \emph{i.e.} $\sigma>\omega_2$ (note $\omega_2>\omega_1$) the work output in all three measurement schemes essentially overlap with the UM value. In contrast, for measurement scheme S3 we see in Fig.~\eqref{fig:ComparisonOfwavg} (c) that even in the limit of $\sigma\to0$ the average work does not approach the TPM value. More specifically, the average work shows oscillations as a function of thermalization time even for very small $\sigma$. This is in line with our discussion in Sec.~\ref{sec:Model} examining the work distribution from scheme S3 in the limiting case of $\sigma \rightarrow 0$.
Moreover this can be traced back to the fact that the steady state of the Otto cycle still retains some coherence even in the limit $\sigma\to0$ for measurement scheme $\mathrm{S3}$. In order to confirm this, in Fig.~\eqref{fig:ComparisonOfDivergences}, we plot the  Kullback-Leibler (KL) divergence  between Gibbs thermal state $\hrho_{\mathrm{th}}=e^{-\beta_c\Hop_c}/Z_c$ with inverse temperature $\beta_c$ and the steady-state corresponding to the $\mathrm{S1}$, $\mathrm{S2}$, $\mathrm{S3}$, UM and TPM as a function of thermalization time $\tau_b$. We see that the KL divergences for the three different schemes show qualitatively the same features as we have seen for the average work output showing the interpolation between TPM and UM for $\mathrm{S1}$ and $\mathrm{S2}$ and the lack of agreement between $\mathrm{S3}$ and TPM even in the small $\sigma$ limit.

Going beyond the first moments, Fig.~\eqref{fig:ProbDist1} shows the work distribution corresponding to the schemes $\mathrm{S1}$, $\mathrm{S2}$ and $\mathrm{S3}$ for different values of the pointer width $\sigma$. The (discrete) work distribution for the TPM scheme is also plotted for comparison. We see that for very large pointer width $\sigma$, the work distributions become a broad Gaussian for all the pointer-based measurement schemes. Though we lose precision in terms of work output in this limit, we still capture the unmonitored average work. As we decrease pointer width $\sigma$, we see that the work distributions start to approach the TPM limit for the schemes S1 and S2. In contrast, as we depict in Fig.~\eqref{fig:ProbDist1} (c) the distribution corresponding to S3 does not approach the TPM one in the $\sigma \to 0$ limit. To see this more clearly we construct the (discrete) work distribution for the scheme S3 with $\sigma = 0$ and plot its difference from the TPM distribution in Fig.~\eqref{fig:ProbDist2}. 

Finally, in order to demonstrate that our results are valid for more general work protocols than the one in Eqs.~\eqref{TLSHamiltoniancomp} \eqref{TLSHamiltonianexp}, in Fig.~\eqref{fig:avgworkGenUnitary} we plot the work output in the different schemes (in units of the TPM work) for the generic unitaries given in Eqs.~\eqref{TLSUnitary1} and \eqref{TLSUnitary2}. We notice two features reinforcing our previous conclusions. Firstly, as the work strokes are made more non-quasistatic ($r$ increased), leading to coherence generation the pointer-based schemes differ from the TPM scheme. While the schemes S1 and S2 tend to the TPM in the small $\sigma$ limit, S3 does not. Finally, coming to the fluctuations of work, in Fig.~\eqref{fig:workfluctuationsGenUnitary} we plot the variance of work (subtracting the TPM variance for comparison) for different schemes. In agreement with the discussion for the perfectly thermalizing cooling stroke, we see that for large pointer width $\sigma$ the variance in all the pointer-based schemes always remains greater than the TPM values. Moreover, it is largest for $\mathrm{S1}$ and smallest for $\mathrm{S3}$. In contrast for the limit of $\sigma\to0$, while the variances for the schemes $\mathrm{S1}$ and $\mathrm{S2}$ approach the TPM value, the one for $\mathrm{S3}$ can take values less than TPM values.

\section{\label{sec:Conclusion}Conclusion}
Coherence is an important quantum resource that can affect the performance of quantum heat engines in a positive manner in certain settings. The standard way of assessing statistics of work and heat via TPM with projective energy measurements destroys coherence and hence cannot take advantage of this resource. One way to mitigate this issue is to replace the projective measurement by weak energy measurement. Along this line, in this paper, we have considered the work and heat statistics in the steady state operation for a finite time Otto cycle with three pointer-based measurement schemes with varying degrees of measurement back-action. 

Using formal analytical expressions of the work and heat statistics for arbitrary working system as well as analytical and numerical results for a TLS working substance we have shown that all of the measurement schemes produce the same average work as the unmonitored case for weak measurement strengths. In contrast in one of the measurement schemes where only two pointers are used (scheme S3), unlike the other two schemes, does not give results that approach the TPM work statistics in the limit of infinitely precise initial pointer states \cite{Son2021}. Moreover we find that by the choice of the measurement scheme as well as the strength of the measurement (characterized by the initial pointer state imprecision) we can control the extent of coherence retention in the steady state of the working substance. Thus, our results demonstrate an ability to control the coherence and hence the work statistics in quantum Otto heat engine cycle using measurements extending and complementing the results in \cite{Son2021,Son2022}. Our study also leads to some questions for future work. For instance, while we have shown that in regimes where the unmonitored work is larger than the TPM, by making the pointer-measurement increasingly weak the average work can approach the unmonitored value. Unfortunately, in this limit the variance of the work diverges. It would be interesting to identify measurement protocols where one can approach the unmonitored work with a finite amount of variance. 

\begin{acknowledgments}
This work was supported and enabled by an MoE
(India) funded Ph.D. fellowship at IIT Gandhinagar (R.S.)
\end{acknowledgments}
\appendix

 \bibliography{mybib3}% Produces the bibliography via BibTeX.
\end{document}